# Field-induced oscillation of magnetization blocking in holmium metallacrown magnet

Si-Guo Wu[1], Ze-Yu Ruan[1], Guo-Zhang Huang[1], Jie-Yu Zheng[1], Veacheslav Vieru[2,7], Gheorghe Taran[3], Jin Wang[1], Yan-Cong Chen[1], Jun-Liang Liu[1]*, Le Tuan Anh Ho[2,4]*, Liviu F. Chibotaru[2]*, Wolfgang Wernsdorfer[3,5,6]*, Xiao-Ming Chen[1], Ming-Liang Tong[1]*

Single-molecule magnets (SMMs) are promising elements for quantum informatics. In the presence of strong magnetic anisotropy, they exhibit magnetization blocking - a magnetic memory effect at the level of a single molecule[1]. Recent studies have shown that the SMM performance scales with the height of magnetization blocking barrier. By employing molecular engineering this can be significantly modified[2,3], remaining independent from other external factors such as magnetic field. Taking advantage of hyperfine coupling of electronic and nuclear spins further enhances their functionality,[4-7] however, a poor understanding of relaxation mechanisms in such SMMs limits the exploitation of nuclear-spin molecular qubits. Here we report the opening discovery of field-dependent oscillation of the magnetization blocking barrier in a new holmium metallacrown magnet driven by the switch of relaxation mechanisms involving hyperfine interaction. Single-crystal magnetic hysteresis measurements combined with first-principles calculations reveal an activated temperature dependence of magnetic relaxation dominated either by incoherent quantum tunneling of magnetization at anti-crossing points of exchange-hyperfine states or by Orbach-like processes at crossing points. We demonstrate that these relaxation mechanisms can be consecutively switched on and off by increasing the external field, which paves a way for manipulating the magnetization dynamics of SMMs using hyperfine interaction.

To meet the demand of explosive growth in information technology, advanced materials towards high-density data storage and quantum information processing have been intensively investigated. Single-molecule magnets (SMMs) in which individual nuclear spin can be embedded controllably and reproducibly[1], are desirable alternatives varying from magneto-memory materials[2] to spin qubit devices[3]. Making good use of nuclear spins, the implementation of quantum Grover's search algorithm[4], electronic read-out and manipulation[5,6], and coherence enhancement via atomic clock transitions[7] have been achieved recently on SMMs. Although quantum relaxation of magnetization has been explored in many systems[8-13], it is still ambiguous for nuclear-spin-driven dynamics in SMMs, leading to the limitation of their further exploitation.

The hyperfine interaction of electronic and nuclear spins introduces new complexity in the magnetic relaxation compared to pure-spin SMMs. The most involved phenomenon is the quantum tunneling of magnetization (QTM) responsible for the demagnetization of SMMs at low temperature[14-18]. It is mediated by transverse crystal field[19], dipole-dipole interactions[20] and dipolar coupling to the nuclear spins of surrounding atoms[21]. Besides, tunneling involving spin-phonon modulation, the incoherent QTM (iQTM)[21], is still elusive in the description of magnetic relaxation transitions between exchange-hyperfine states of polynuclear SMMs. At a higher temperature, additional relaxation mechanisms come into play (direct, Orbach and Raman processes[1]) which in the presence of hyperfine coupling complicate significantly the magnetization dynamics. As a result, a decent microscopic description of magnetic relaxation in such systems has never been attempted. It becomes clear, that further progress in the exploration of nuclear-spin-driven relaxation mechanisms requires a detailed microscopic modelization.

Here we present a complex study of polynuclear SMM exhibiting nuclear-spin-driven magnetic relaxation. We employed $^{165}$Ho$^{III}$ (nuclear spin $I$ = 7/2) as electron- and nuclear-spin carrier owing to its strong magnetic anisotropy and simplex natural abundance of isotopes. Given Ho$^{III}$ is a non-Kramers ion, in order to avoid the splitting of its spin-orbit doublets, the combination of symmetry strategy[22] and metallacrown (MC) approach[23,24] were came up with and afforded a crown-like complex Ho$^{III}$F$_2$[15–MC$_{Ni}$–5] (**1**) with the axially-ligated F–Ln–F motif. Thanks to the resulting pentagonal-bipyramidal geometry ($D_{5h}$), the transverse magnetic anisotropy is minimized,[25,26] while the axial one is strongly enhanced.[27] More importantly, the interaction of the nuclear spin of Ho$^{III}$ with the exchange states of the {Ni$^{II}_5$} MC ring provides a magnetic complexity which leads to temperature- and field-dependent magnetization relaxation behavior not observed before and fully elucidated here.

[1]Key Laboratory of Bioinorganic and Synthetic Chemistry of Ministry of Education, School of Chemistry, Sun Yat-Sen University, 510275 Guangzhou, Guangdong, P. R. China. [2]Theory of Nanomaterials Group, Katholieke Universiteit Leuven, Celestijnenlaan 200F, 3001 Leuven, Belgium. [3]Institute of Nanotechnology (INT), Karlsruhe Institute of Technology (KIT), Hermann-von-Helmholtz-Platz 1, D-76334, Eggenstein-Leopoldshafen, Germany. [4]Department of Chemistry, National University of Singapore, 3 Science Drive 3 Singapore 117543. [5]Institut Néel, CNRS and University Grenoble-Alpes, 25 Rue des Martyrs, F-38042, Grenoble, France. [6]Physikalisches Institut, Karlsruhe Institute of Technology, 76131 Karlsruhe, Germany. [7]Faculty of Science and Engineering, Maastricht University, Lenculenstraat 14, 6211 KR Maastricht, The Netherlands.

liujliang5@mail.sysu.edu.cn; chmhlta@nus.edu.sg; Liviu.Chibotaru@kuleuven.be; wolfgang.wernsdorfer@kit.edu; tongml@mail.sysu.edu.cn

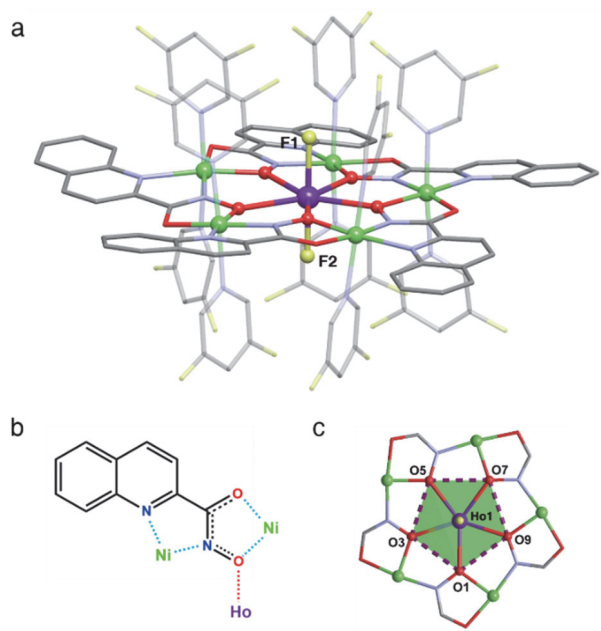

**Fig. 1 | Structure of Ho(III)F$_2$[15−MC$_{Ni}$−5] (HoNi$_5$). a**, Molecular structure of HoNi$_5$. The anion, solvent molecules and hydrogen atoms are omitted for clarity. Color code: purple (Ho), green (Ni), grey (C), red (O), light blue (N) and yellow (F). **b**, The [3.2111] coordination mode of quinha$^{2-}$ ligand. **c**, Top view of [15−MC$_{Ni}$−5] metallacrown.

Reaction of metal salts and quinaldichydroxamic acid (H$_2$quinha) with the excess of ammonium fluoride as well as 3,5-difluoropyridine (dfpy) afforded two charming 3d-4f metallacrowns, [Ln$^{III}$Ni$^{II}_5$(quinha)$_5$F$_2$(dfpy)$_{10}$](ClO$_4$)·2EtOH (Ln = Ho, **1**; Y, **2**). Single-crystal X-ray diffraction reveals that these compounds are isostructural and crystallize in the triclinic space group *P*-1 (Table S1). In complex **1**, the Ho$^{III}$ ion is surrounded equatorially by a [15−MC$_{Ni}$−5] ring while axially encapsulated with two terminal fluoride ions. (Fig. 1a) The {HoF$_2$} moiety is protected from bridging or hydrogen bonding through F$^-$ thanks to the steric hindrance of the bulky 3,5-difluoropyridine connecting with Ni$^{II}$ ions axially on both sides. The tetradentate ligand quinha$^{2-}$, employing a [3.2111] coordination mode[28], chelates the adjacent Ni$^{II}$ ions with two nitrogen atoms and two oxygen atoms, respectively. Five such subunits self-assembly in a head-to-tail fashion giving rise to a neutral [15−MC$_{Ni}$−5] ring. In this case, Ho$^{III}$ ion is captured in the cavity provided by 5 hydroximate oxygen atoms (Fig. 1c). Consequently, the Ho$^{III}$ site possesses a compressed pseudo $D_{5h}$ coordination environment with the mean Ho−O and Ho−F bond lengths of 2.446(2) Å and 2.123(2) Å, respectively. In addition, the equatorial O−Ho−O angles range from 71.17(6) to 72.36(6)° and the axial F−Ho−F angle lines up at 176.56(7)°, approaching the $D_{5h}$ geometry. All the Ni$^{II}$ ions are 6-coordinate with a [N$_4$O$_2$] octahedral geometry and the distances between the neighbor Ni$^{II}$ sites vary within 4.673−4.716 Å. The lanthanide ions are well-separated by the MC rings with the nearest Ho···Ho distance of 13.124 Å.

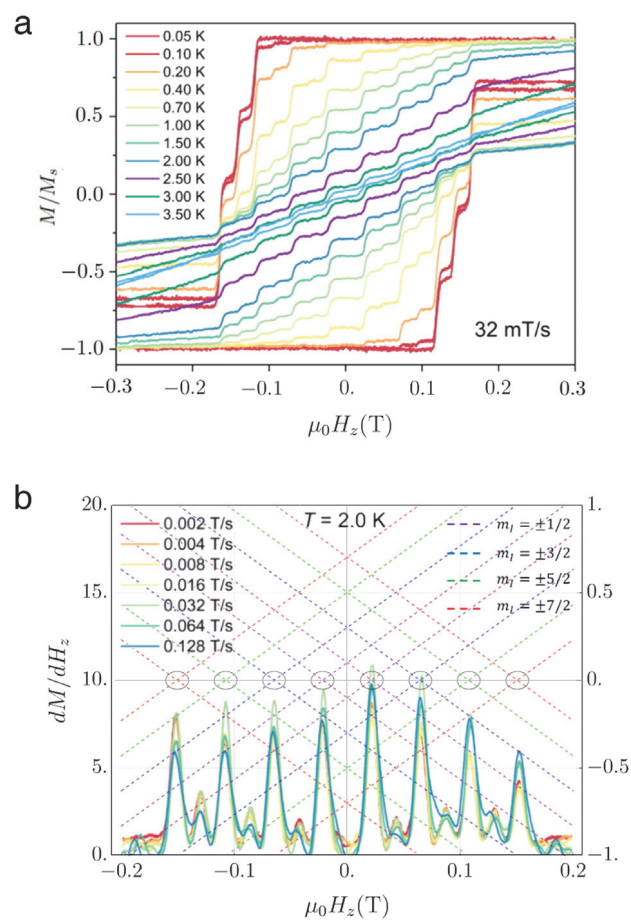

**Fig. 2 | Single-crystal magnetic hysteresis loops performed on micro-SQUID apparatus. a**, Magnetic hysteresis loops (0.05–3.5 K) obtained with a sweep rate of 32 mT s$^{-1}$ for the yttrium diluted sample of **1** with Ho:Y = 1:19. **b**, The 1$^{st}$ derivatives of magnetization with variable sweep rates (2–128 mT s$^{-1}$) at 2.0 K and the Zeeman energy diagram (along the easy axis) for the low-lying 16 hyperfine states drawn for the hyperfine coupling constant $A_{hf}$ = 0.0251 cm$^{-1}$. The black circles, matching well the maxima of 1$^{st}$ derivative of magnetization for single-crystal sample, correspond to the anti-crossings of hyperfine levels along easy axis.

The open magnetic hysteresis loops, using a sweep rate of 0.02 T s$^{-1}$ for **1** (Fig. S12), are observed below 5 K. On lowering the temperature, the coercivity increases to 0.145 T at 1.8 K. Moreover, staircase-like steps are monitored within *ca.* ±0.2 T at 1.8 K upon decreasing the sweep rate (Fig. S13). These steps reveal the appearance of magnetization reversals. To gain further insight, we investigated the low-temperature magnetic properties of the yttrium diluted sample with Ho:Y = 1:19 by performing micro-SQUID measurements under different field sweep rates ($dB_z/dt$) on a single crystal (0.05–3.5 K). Fig. 2a shows the temperature dependence of the hysteresis loops at a fixed sweep rate of 32 mT s$^{-1}$. A hysteresis up to 3.5 K bearing evidence of large uniaxial anisotropy is observed. Moreover, the existence of multiple steps in the hysteresis suggests a well resolved energy structure likely resulting from hyperfine interaction on Ho$^{III}$ (Fig. 3a). At very low temperatures, sweeping the field from the saturation point to zero starting, we observe that the molecules relax mainly at its specific values. This can be explained by the anti-crossing of mixed states of hyperfine origin, which accordingly open tunneling channels towards states with opposite magnetic moment (Fig. 3b, c). The hysteresis steps reflect both the magnitude of the tunnel splitting and the population of the hyperfine states. Thus, at very low temperature ($T$ = 0.05–0.2 K), where only the lowest exchange-hyperfine energy levels are populated, the relaxation only occurs when the external field brings these states into resonance (Fig. 3c).

Important features of the relaxation process can also be observed in the derivatives of the hysteresis curves (Fig. 2b). Thus equally spaced transitions around zero field seen as sharp peaks in $dM/dB_z$ reveal the existence of repeating switches between slow and fast population transitions. According to our model, these peaks correspond to the hyperfine crossings that conserve the nuclear spin and reverse the electron spin. Additionally, positions of the peaks at ±22, ±65, ±108, ±151 mT, with the period of $\Delta B$ = 43 mT, lead to a theoretical value of the hyperfine constant $A_{hf}$ = $g\mu_B\Delta B$ = 0.0251 cm$^{-1}$. The parity effect occurring around $H_{dc}$ = $kH_0$ ($k$ = −7, −6, ... , 7; $H_0$ = 22 mT) (Figs. 3c and S71) originates from the resonance of hyperfine states. At even $k$ hyperfine states with the same nuclear spin quantum number ($\Delta m_I$ = 0) are totally out of resonance. Yet some hyperfine states with $\Delta m_I \neq 0$ can also exhibit at resonance slight QTM due to the transversal magnetic field, manifested as small peaks in the derivative of magnetization (Fig. 2b).

The relaxation time ($\tau$) of **1** at 8 K was investigated through the alternating-current (ac) magnetic susceptibility (Fig. 4a). $\tau(H)$ clearly displays oscillations in function of applied field, with four minima located at 30, 80, 130 and 170 mT matching the peak positions of $dM/dB_z$ for polycrystalline sample (Fig. S31) but spaced larger than in single crystal (Fig. S15).

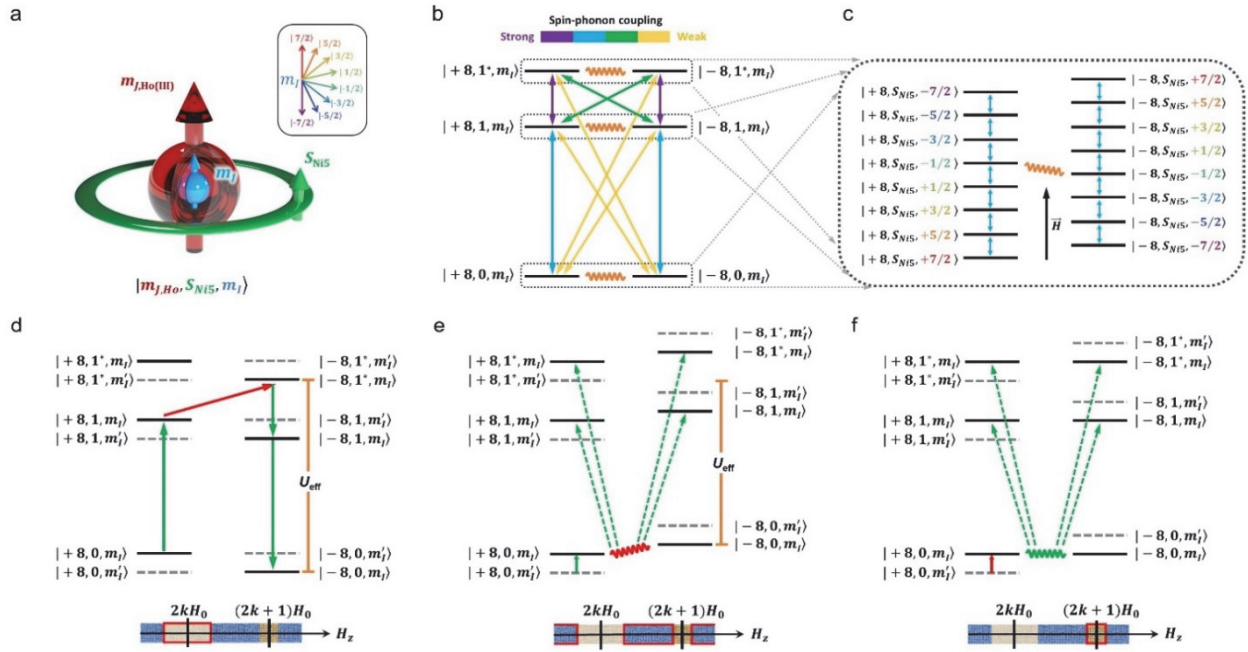

**Fig. 3| Hyperfine interaction and dominant relaxation paths/regimes in HoNi$_5$ molecule at low temperature. a,** Three coupled spin components in HoNi$_5$: eight-level nuclear spin ($m_I$) of Ho$^{III}$ ion is hyperfine coupled to the electronic spin projections ($m_{J,Ho}$) which, at they turn, couple ferromagnetically to $S_{Ni5}$ of the {Ni$_5$}. **b, c,** Possible transition paths in HoNi$_5$ at low temperature between exchange states and within an exchange doublet. **d-f,** Dominant relaxation processes in **1** at low temperature in external field when hyperfine-split states are respectively in the proximity of or right at the crossing points (1$^{st}$ relaxation regime); between the crossing and anti-crossing points (2$^{nd}$ relaxation regime); very close or right at the anti-crossing points (3$^{rd}$ relaxation regime). The red frames in lowest plots indicate the domain of $H_z$ (applied along the anisotropy axis) at which the corresponding mechanism is dominant. Solid/dashed/wavy lines stand for the direct/escape electron-phonon processes/quantum tunneling of magnetization; green lines are dominant (fast) and the red lines are bottleneck (slow) relaxation steps determining the temperature dependence of the transition rate in each process.

To elucidate the origin of relaxation in **1**, a minimal microscopic model was built based on *ab initio* calculations (see the Supporting Information). Thus the seven lowest exchange doublets (Table S9 and Scheme S2) were simulated with three doublets, each belonging to $S_{Ni5}$ = 0, 1 and 1*, respectively, (Fig. 3b) with averaged energy values and tunneling splitting gaps (Table S10). These states are hyperfine split and marked as $|\pm 8,0,m_I\rangle$, $|\pm 8,1,m_I\rangle$ and $|\pm 8,1^*,m_I\rangle$, respectively (Fig. 3b). In an applied field, a crossing or anti-crossing of the hyperfine components in each of these multiplets may occur, resulting in three distinct relaxation regimes (Figs. 3d-f). In particular, in the proximity of or right at the crossing points, ($H_z = 2kH_0$, $H_0 = A_{hf}/2g\mu_B$, $k$ = −3... 3), as tunneling transitions are suppressed, relaxation of the system takes place via the slow Orbach-like process (1$^{st}$ relaxation regime, Fig. 3d). As the system moves away from the crossing points, the quantum tunneling process becomes stronger. At some magnetic field, it dominates over the Orbach-like process. The relaxation in the system is then mainly governed by the incoherent quantum tunneling of magnetization (2$^{nd}$ relaxation regime, Fig. 3e). Moving closer to anti-crossing points ($H_z = (2k+1)H_0$, $k = -4...3$), the resonance between the exchange-hyperfine states $|\pm 8, S_{Ni}, m_I = -(2k+1)/2\rangle$ fully opens corresponding tunneling channel, thus allowing for a very fast iQTM between these states. Accordingly, the relaxation of the system is determined by the rate of direct transition between hyperfine states within each exchange multiplet (3$^{rd}$ relaxation regime, Fig. 3f). Repetition of this scenario with the increase of applied field leads to the observed oscillation of the relaxation time and the effective blocking barrier (see below). Within our model the oscillations in $\tau(H)$ are reproduced very well with a slightly larger hyperfine coupling constant ($A_{hf}$ = 0.0270 cm$^{-1}$) due to polycrystallinity of the sample. Another noticeable feature is the damping of the oscillation of $\tau(H)$ (Fig. 4a). This is mainly the result of the existence of different relaxation regimes and the distribution of relaxation times due to the random orientation of the molecules (see the Supporting Information for details). Thus in Fig. 4a, while the fraction of molecules in the slow 1$^{st}$ relaxation regime decreases with the field, the one in the faster 2$^{nd}$ and 3$^{rd}$ relaxation regimes increases concurrently (Fig. S79). These variations are especially drastic from the 1$^{st}$ ($H = 0$) to the 2$^{nd}$ ($H \approx 55$ mT) peak. Accordingly, the observed relaxation times corresponding to these peaks show a decelerating damping from peak to peak under increasing field. The same reasoning can also be applied for the minima at the oscillating curve in Fig. 4a. In this case the percentage of molecules in the 1st relaxation regime keeps rising, thus increasing the relaxation time at the minima. However at large $H$ the enlargement of the tunneling splitting gap due to transversal field components ($\sim H_{tr}^4$) increases the effect of the 2$^{nd}$ relaxation regime and compensates this magnitude, resulting in slightly lower values for the 3$^{rd}$ and 4$^{th}$ minima.

Under applied dc fields of 0–0.2 T, the ac out-of-phase susceptibility of **1** exhibits a peak up to 50 K which shifts back and forth in frequency with a progressively narrowing amplitude as the dc field increases (Fig. S70). To get further insight, the relaxation time at different temperatures was extracted from ac data by using a generalized Debye model. The plot of relaxation time versus temperature (Fig. 4b) reveals two linear regions (4–36 K and 38–50.5 K). $\tau$ values proved practically independent from applied field in the high temperature region. Yet at lower temperature the values oscillate significantly with the field. To our knowledge this behavior was not observed before in SMMs.

The plot in Fig. 4b, fitted with equation $\tau^{-1} = \tau_{0(1)}^{-1}\exp(-U_{eff}/k_BT) + \tau_{0(2)}^{-1}\exp(-\Delta/k_BT)$, gives a shared $U_{eff}$ = 577(6) cm$^{-1}$ and $\tau_{0(1)}$ = 3.3(5)–3.6(6) × 10$^{-13}$ s while $\Delta$ and $\tau_{0(2)}$ vary within 18.6(1)–22.9(1) cm$^{-1}$ and 5.7(1)–9.4(2) × 10$^{-4}$ s, respectively. The fact that there is only one $U_{eff}$ for the temperature-dependent relaxation in the high temperature domain indicates that the relaxation in this regime mainly comes from the Ho$^{III}$ ion, which is confirmed by *ab initio* calculations showing in addition that $U_{eff}$ corresponds to the 2$^{nd}$ excited doublet (Fig. S75). Remarkably, our microscopic model reproduces very well the oscillations of $\Delta$ for varying $H$ (Fig. 4c).

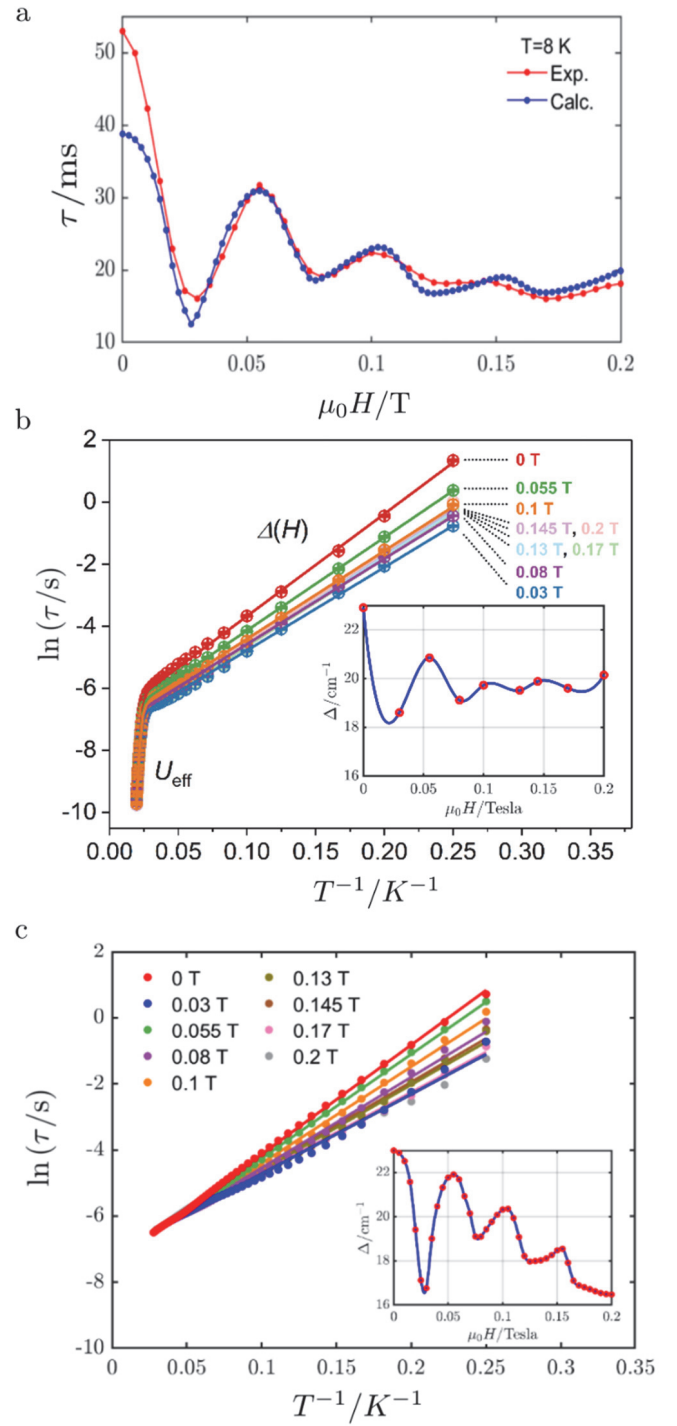

**Fig. 4| Relaxation dynamics for polycrystalline sample of 1. a**, Field dependence of $\tau$ extracted from ac susceptibility measurements (red cycles) and from theory (blue line). **b,** Temperature dependence of $\tau$ extracted from ac susceptibility measurements (circles). Solid lines are the best fits with the equation: $\tau^{-1} = \tau_{0(1)}^{-1}\exp(-U_{eff}/k_BT) + \tau_{0(2)}^{-1}\exp(-\Delta/k_BT)$ with the same $U_{eff}$ ($R^2$ = 0.99918). **c**, Temperature dependence of $\tau$ from theoretical simulations (circles). Lines are the best fits with the equation $\tau^{-1} = \tau_{0(2)}^{-1}\exp(-\Delta/k_BT)$ in the temperature range 4–36 K. **Insets in (b) and (c):** field dependence of the activation energy (magnetization blocking barrier) from experiment and simulation, respectively; solid lines are the guide for the eyes.

As shown in the insets in Figs. 4b,c, $\Delta$ displays oscillating behavior with the same period and similar damping as $\tau(H)$. This oscillation and the exponential temperature-dependence of the relaxation time can again be explained by the consecutive switch between several relaxation regimes (Figs. 3d-f) with increasing the field. Thus, the peaks of $\Delta(H)$ correspond to the domains around the crossing points where the Orbach-like process dominates (Fig. 3d). In this activated process, the activation energy will have a value close to the energy of the 2$^{nd}$ excited exchange doublet $|\pm 8,1^*,m_I\rangle$, 24.4 cm$^{-1}$ according to *ab initio* calculations. This value will decrease with increasing the field mainly due to the polycrystallinity of the sample, very much as the damping of the $\tau(H)$ oscillation. Meanwhile, the minima of the $\tau(H)$ curve correspond to those domains around the anti-crossing points where majority of the molecules are in the 2$^{nd}$ relaxation regime (Fig. 3e). The activation character of the relaxation time originates in this case from the average escape rate of the iQTM process[1] from the ground exchange doublet to the 1$^{st}$ (18.4 cm$^{-1}$ from *ab initio* calculations) and 2$^{nd}$ excited ones, which is lower than the one from the Orbach-like process. The increase of the minimal values of the activation energy is also due to the the polycrystallinity of the sample. Despite some deviations in the damping of the blocking barrier (Figs. 4b,c), the microscopic model gives a good description of the oscillation of magnetization blocking barrier.

The observed field-dependent oscillatory behaviour requires the presence of strong hyperfine interaction at the magnetic center. Its exchange coupling with surrounding metal ions builds up a spectrum of mixed states for which several relaxation mechanisms can be periodically switched on and off under varying magnetic field. This finding opens new perspectives for the manipulation of magnetic relaxation in complex SMMs via hyperfine interaction.

**Online content**

Any methods, additional references, supplementary information, acknowledgements, author contributions and competing interests are available at https://doi.org/...

**Data availability**

All the data and simulation are available in the paper and Supplementary Information, and/or from the corresponding authors upon reasonable request.

**Code availability**

Fortran source codes are available in the Supplementary Information, and/or from the corresponding authors upon request.

**Acknowledgements**

This work was supported by the National Key Research and Development Program of China (2018YFA0306001), the NSFC (Grant nos 21620102002, 21822508, 21701198, 21805313 and 21821003), and the Pearl River Talent Plan of Guangdong (2017BT01C161). V.V. acknowledges the postdoctoral fellowship of Fonds Wetenschappelijk Onderzoek Vlaanderen (FWO, Flemish Science Foundation). L. T. A. H. acknowledges the research grants R-143-000-A65-133 and R-143-000-A80-114 of the National University of Singapore.

**Author contributions**

Ming-Liang Tong conceived and designed the research project. Si-Guo Wu, Ze-Yu Ruan, Jie-Yu Zheng and Jin Wang executed the synthesis and finished the crystallographic analysis. Guo-Zhang Huang, Yan-Cong Chen performed and analyzed the dc and ac magnetization measurements for crystalline samples. Gheorghe Taran and Wolfgang Wernsdorfer performed and analyzed the single-crystal micro-SQUID magnetization measurement. Si-Guo Wu performed and analyzed the PXRD, EA and IR measurements. Veacheslav Vieru, and Liviu F. Chibotaru carried out the ab initio calculations. Le Tuan Anh Ho and Liviu F. Chibotaru designed the theoretical models, performed the simulations and explained the relaxation mechanisms. Si-Guo Wu, Ze-Yu Ruan, Gheorghe Taran, Jun-Liang Liu, Wolfgang Wernsdorfer, Le Tuan Anh Ho, Liviu F. Chibotaru, Xiao-Ming Chen and Ming-Liang Tong were involved in writing the manuscript and they have all given their consent to this publication.

**Competing interests**

The authors declare no competing interests.